# Virtual Telescope for X-Ray Observations

Kyle Rankin, Steven Stochaj
New Mexico State University
Las Cruces NM 88003; (575) 646-3115
krankii@nmsu.edu

John Krizmanic
Center for Space Science & Technology – Goddard Space Flight Center
University of Maryland, Baltimore County
Greenbelt MD, 20771; (301) 286-6817
John.F.Krizmanic@nasa.gov

Neerav Shah
Goddard Space Flight Center
Greenbelt MD, 20771; (301) 286-8174
neerav.shah-1@nasa.gov

Asal Naseri
Space Dynamics Laboratory
Logan UT, 84341; (435) 713-3400
asal.naseri@sdl.usu.edu

**ABSTRACT**

Selected by NASA for an Astrophysics Science SmallSat study, The Virtual Telescope for X-Ray Observations (VTXO) is a small satellite mission being developed by NASA's Goddard Space Flight Center (GSFC) and New Mexico State University (NMSU). VTXO will perform X-ray observations with an angular resolution around 50 milli-arcseconds, an order of magnitude better than is achievable by current state of the art X-ray telescopes. VTXO's fine angular resolution enables measuring the environments closer to the central engines in compact X-ray sources. This resolution will be achieved by the use of Phased Fresnel Lenses (PFLs) optics which provide near diffraction-limited imaging in the X-ray band. However, PFLs require long focal lengths in order to realize their imaging performance, for VTXO this dictates that the telescope's optics and the camera will have a separation of 1 km. As it is not realistic to build a structure this large in space, the solution being adapted for VTXO is to place the camera, and the optics on two separate spacecraft and fly them in formation with the necessary spacing. This requires centimeter level control, and sub-millimeter level knowledge of the two spacecraft's relative transverse position. This paper will present VTXO's current baseline, with particular emphasis on the mission's flight dynamics design.

## INTRODUCTION

The Virtual Telescope for X-Ray Observations (VTXO) is a new Phase Fresnel Lens (PFL) based X-ray telescope which will perform X-ray observations with an angular resolution around 50 milli-arcseconds. This is an order of magnitude improvement over Chandra, the current state of the art X-ray telescope. PFL optics can generate nearly diffraction-limited imaging using current manufacturing technologies, such as MEMS. However, in order to do so the PFL requires an extremely long focal length on the order of 1 km; even with the modest few centimeter lenses being used on VTXO. As this design can't be achieved with traditional rigid telescope structures, the VTXO mission will use a formation flying scheme that has been developed where two spacecraft, one with PFL optics and the second with an X-ray camera are flown in a formation which approximates a rigid structure. The VTXO mission has recently undergone a Mission Planning Lab (MPL) at NASA's Wallops Flight Facility. Development work on the VTXO mission is ongoing and will result in a mission proposal in the near future. Beyond VTXO, the formation flying technology being developed for the mission should be expandable to be used for future precision formation flying missions, such as a distributed aperture telescope.

## PHASE FRESNEL LENS

Phase Fresnel Lenses (PFLs) provides an opportunity to exceed the performance in imaging preformance of conventional Wolter type-1 X-ray optics used in X-ray telescopes such as Chandra by utilizing an entirely new form of optics.3 PFLs work on principles of diffraction, effectively modifying the phase of the light into the lenses focal point.4,5 In theory this results in an lens with 100% efficiency. In testing, a prototype lens achived near diffraction-limited imaging in the X-ray with efficiencies of ~30% or better A 3 cm diameter PFL is

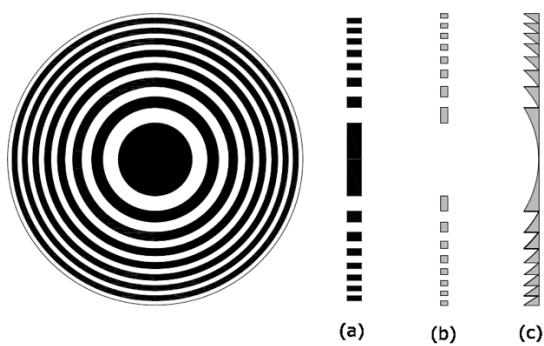

Figure 1 (a) Fresnel Zone Plate, (b) Phase Zone Plate, (c) Phase Fresnel Lens.2

estimated to have an X-ray sensitivity to brighter X-ray sources in the imaging band around 4.5 keV.8,9

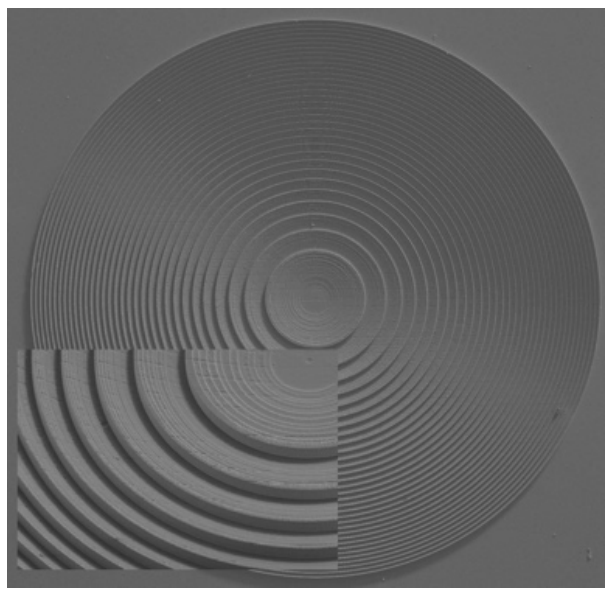

Figure 2: SEM image of a prototype PFL.8

## SCIENCE MISSION

X-ray observations are a crucial part of astronomical observations, permitting imaging of some of the most energetic phenomena in the universe. The location of these environments are associated with massive compact objects such as black holes, neutron stats, and super novae. Furthermore, X-rays are often associated with stellar flares, both on the surface of the sun and other stars.

With its modest size lens, VTXO is limited to imaging relatively bright sources. Nonetheless given VTXO's order of magnitude improvement in angular resolution compared to what has been achieved to date in X-ray measurements, numerous targets within VTXO's capabilities have been identified which provide a substantial scientific return. Table 1 lists the preliminary targets that have been selected for the VTXO mission.

**Table 1: Table 1: Baseline ability of VTXO to collect 1000 photons from bright compact X-ray sources assuming a 150 eV FWHM energy resolution around 45.54 keV and a 3cm diameter PFL with 40% efficacy. $\Gamma$ is the spectral index and the flux is that reported in the energy range of ~2 – 10 keV.**

| Source | $\Gamma$ | Flux (erg/s/cm$^2$) | Collection Time (hr) |
|---|---|---|---|
| Sco X-1 | -2.1 | 2 x 10$^{-7}$ | 0.2 |
| Cyg X-1 Soft | -1.5 | 3 x 10$^{-8}$ | 1.2 |
| Cyg X-1 Hard | -2.1 | 1 x 10$^{-8}$ | 3.4 |
| Cyg X-3 Soft | -1.5 | 2 x 10$^{-8}$ | 1.7 |
| Cyg X-3 Hard | -2.1 | 8 x 10$^{-9}$ | 4.6 |
| GX 5-1 | -4.7 | 5 x 10$^{-8}$ | 1.4 |
| Crab Pulsar | -2.1 | 2 x 10$^{-9}$ | 16 |
| γCas | -1.7 | 3 x 10$^{-10}$ | 115 |

The VTXO science mission is continuing to evolve alongside the VTXO spacecraft as the capabilities of the telescope continue to evolve and be better understood.

## BASELINE SPACECRAFT DESIGN

VTXO is an X-ray telescope consisting of two spacecraft, the Optics Spacecraft (OSC) a 6U CubeSat carrying Phase Fresnel Lenses (PFLs), and the detector spacecraft (DSC) an ESPA-class small-sat with an X-ray camera. These two spacecrafts will fly in a formation approximating a rigid telescope at 1 km separation. During formation flying, the OSC will fly along a natural orbit trajectory, while the DSC will maneuver in a sudo-orbit to maintain a rigid formation with the OSC. The DSC will have a relative navigation sensor based on a star tracker to track a beacon located on the OSC. Range will then be determined by using radio ranging over an intersatellite radio link.

*Detector Space Craft*

The DSC is an ESPA-class spacecraft with an estimated wet mass of ~100 kg. This spacecraft caries a 100 m/s cold gas propulsion system, a star tracker based relative navigation sensor, along with the X-ray camera, and radiation hardened avionics.

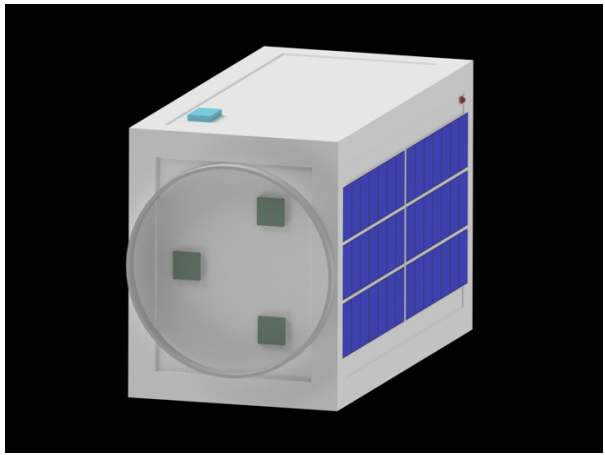

**Figure 3: Image of the DSC**

*Optics Space Craft*

The OSC is a 6U CubeSat, with ~10 kg wet mass. The OSC carries the PFLs, and a cold gas propulsion system to provide an initial perigee raise and provide orbit maintenance maneuvers. The OSC deploys from a standard 6U canister and contains a radiation tolerant avionics system.

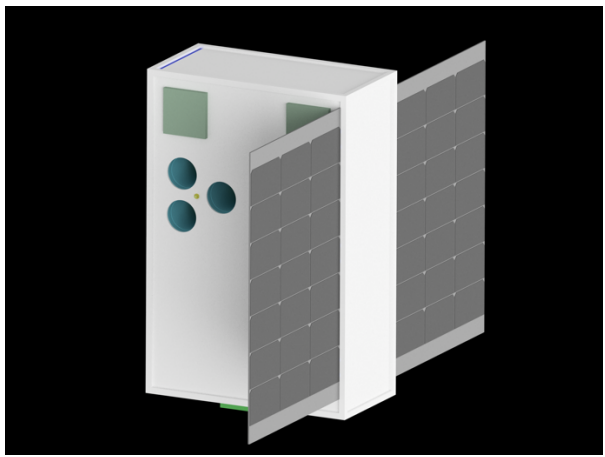

**Figure 4: Image of the OSC**

**CONCEPT OF OPERATIONS**

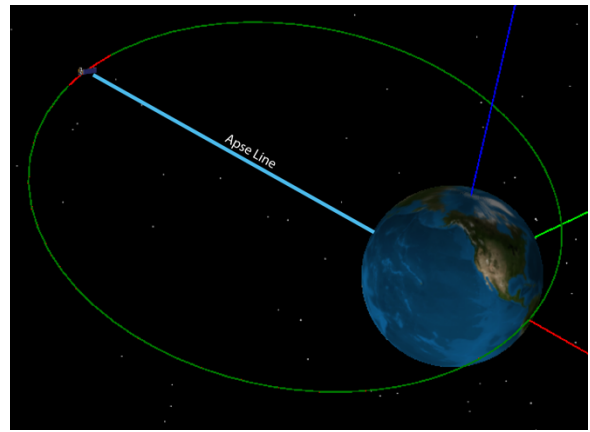

**Figure 5: Representation of VTXO orbit. The red portion is the observation portion of the orbit. Graphic not to scale.**

The two spacecraft launch on a common rideshare to a highly elliptical super synchronous transfer orbit. Once deployed, the they will perform a checkout procedure after which the two spacecraft will perform a series of maneuvers to bring the two spacecraft back together. At that point, VTXO is ready to begin performing the science phase of the mission. During the science phase, all observations will be performed near apogee where the relative gravity gradient is minimal which minimizes fuel consumption. As the vehicles approach perigee, they will be brought into a close formation with around 20 m separation which will reduce fuel consumption as they pass through perigee. After going through perigee, the two vehicles will then be maneuvered back out to the 1 km separation for the observation portion of the orbit. The observation portion of the orbit will be conducted for approximately ±5 h on either side of apogee, giving 10 hours of observation time during each orbit.

**RELATIVE NAVIGATION SENSORS**

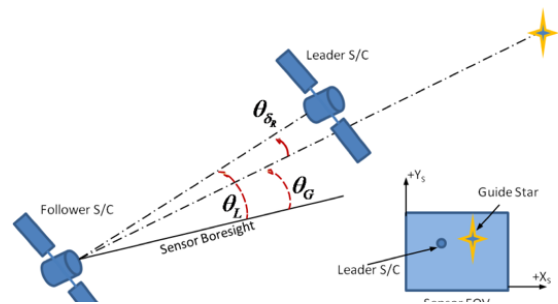

**Figure 6: Description of relative navigation sensor. 17**

The VTXO navigation system is based around three sensors, a radio crosslink with ranging, a GPS, and a star tracker based relative navigation camera. The basis of the navigation system is the NISTEX-II star tracker mounted on the DSC. This camera is capable of determining the bearing to the OSC with accuracies of better than 100 milli-arcseconds. The NISTEX-II will work as shown in Figure 6 by comparing the location of a beacon on the OSC with the background starfield. This will provide two axis relative position knowledge; the third axis will be provided by a radio ranging system, which provides ranging on the order of a meter.16 In addition to these relative navigation sensors, a GPS system on each spacecraft will be used for orbit determination, and for the reacquisition of the OSC if NISTEX-II loses track.

## CONTROL SYSTEM

VTXO operates in a nearly linear control environment.17 This is a well understood environment, with broad industry experience in linear control systems such as PID controllers dating back decades. Providing an actuator with a sufficiently low minimum impulse bit is used, and adequate position knowledge is available, linear control for this system is relatively simple. Given that the control requirements are nearly an order of magnitude lower than the knowledge requirements on VTXO, it is not anticipated that there will be any problem with position knowledge.

## PROPULSION

The baseline VTXO design calls for cold gas propulsion. While somewhat limiting in terms of total mission life, cold gas has proven flight heritage. The baselined cold gas system is based on the unit flown on JPL's MarCo mission.19 and as such the unit has demonstrated flight heritage. Both mono-prop, and electrospray propulsion continue to be evaluated as alternatives for the mission. Both these options would both provide significant increases in mission life, and in the case of electrospray a substantial reduction in minimum impulse bit, with a corresponding improvement in control precision. However, these advantages will continue to be evaluated relative to the system's TRL level. Table 2 shows the capabilities of VTXO's baseline cold gas propulsion system.

**Table 2: VTXO propulsion system baseline**

|     | Dry Mass | Prop Mass | $I_{sp}$ | Total Delta V |
| --- | --- | --- | --- | --- |
| OSC | 13.06 kg | 1.92 kg | 40 s | 53.8 m/s |
| DSC | 102.5 kg | 33.9 kg | 40 s | 111.9 m/s |

## FORMATION FLIGHT DYNAMICS

VTXO utilizes a leader follower dynamic with the OSC traveling on a natural trajectory, the DSC then flies along a sudo orbit with a constant offset from the OSC. In this sudo orbit, the DSC will continuously use its propulsion system to maintain the desired position relative to the OSC in the relative frame, this is necessary because the DSC is not on a natural orbit trajectory.15 Figure 8 shows the position of the DSC in an ICRF aligned frame centered on the OSC as it moves through the orbit between the observation, and the perigee sections of the orbit.

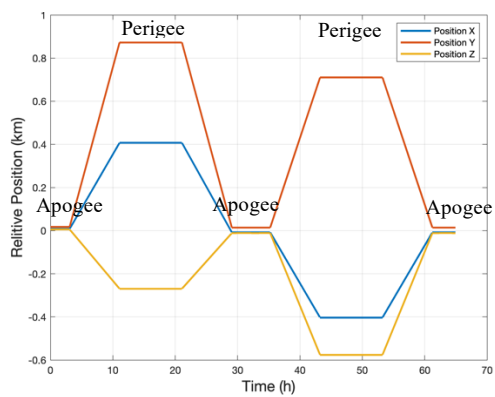

**Figure 7: Plot of the position of the DSC in the relative frame over two orbits**

Figure 7 shows the estimated DSC delta V usage over two orbits using a specific GN&C control algorithm The 2-norm value assumes that the thrusters are pointed in the ideal direction, while the 1 norm value calculates the Delta V for the baseline DSC thruster layout where the thrusters are not necessarily in the optimal orientation.

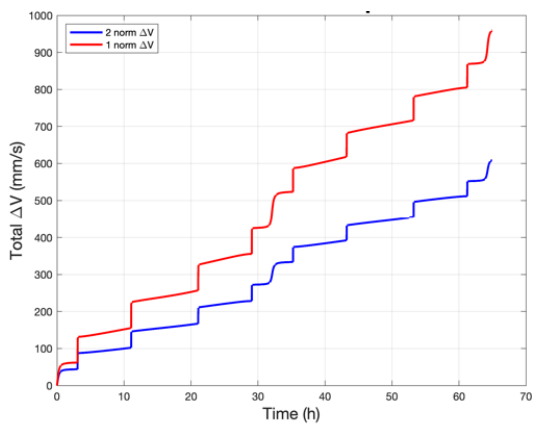

**Figure 8: DSC Delta V over two orbits**

As can be seen in Figure 9 the acceleration required by the propulsion system to maintain formation is orders of magnitude greater as the spacecraft passes through perigee, as compared to apogee. This shows why the DSC is pulled into a close formation with the OSC as the two-spacecraft pass through perigee.

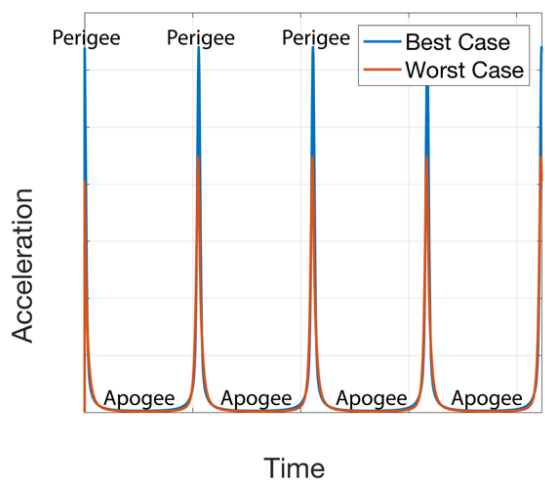

**Figure 9: Acceleration over Time Plot**

Current lifetime estimates are based on the 100 m/s Delta V assuming 0.95 m/s per orbit is approximately 150 days.

**Table 3: VTXO Delta V Budget**

| Maneuver | OSC | DSC |
|---|---|---|
| Attitude Control | 2 m/s (22 N-s calculated value) | 1 m/s (40 N-s calculated value) |
| Perigee Raise | 20 m/s | 10 m/s |
| Orbit Maintenance | 10 m/s | 20 m/s |
| Science Observation | 0 | 65m/s |
| Contingency | 10 m/s | 15 m/s |
| Total | 42 m/s | 111 m/s |

## CONCLUSION

VTXO is a viable mission which is that is capable of numerous scientific observations. The mission is capable of delivering X-ray imaging with more than an order of magnitude improvement in angular resolution over current state of the art X-ray telescopes. This high, resolution imagining gives VTXO the potential to significantly improve the understanding of high-energy phenomenon, and other X-ray sources. Finally, the formation flying technology being developed for VTXO should be expandable to enable future distributed aperture telescope missions, a key technology for future high angular resolution astronomical observations.